\documentclass[twocolumn,nofootinbib, amsmath,amssymb,aps,prd,]{revtex4-1}
\usepackage{CJKutf8}
\usepackage{graphicx}
\usepackage{dcolumn}
\usepackage{grffile}
\usepackage{tikz}
\usepackage{bm}
\usepackage{slashed}
\usepackage{physics}
\usepackage[colorlinks=true,linkcolor=blue,citecolor=blue, urlcolor=blue]{hyperref}
\usepackage{xcolor}

\usepackage{asymptote}

\def\6{\partial}

\newcommand{\be}{\begin{equation}}
\newcommand{\ee}{\end{equation}}
\newcommand{\bea}{\begin{eqnarray}}
\newcommand{\eea}{\end{eqnarray}}

\begin{document}
\preprint{CERN-TH-2021-106}

\title{Bremsstrahlung photons from stopping in heavy-ion collisions}

\author{Sohyun Park}
\email{sohyun.park@cern.ch}
\author{Urs Achim Wiedemann}
\email{urs.wiedemann@cern.ch}
\affiliation{Theoretical Physics Department, CERN, CH-1211 Gen\`eve 23, Switzerland}

\date{\today}

\begin{abstract}
We examine the spectrum of bremsstrahlung photons that results from the stopping of the initial net charge distributions in ultrarelativistic nucleus-nucleus collisions at the CERN Large Hadron Collier (LHC). This effect has escaped detection so far since it 
becomes sizable only at very low transverse momentum and at sufficiently forward rapidity. We argue that it may be within 
reach of the next-generation LHC heavy-ion detector ALICE-3 that is currently under study, and we comment on the physics
motivation for measuring it.
\end{abstract}

\maketitle

\section{Introduction}
In ultrarelativistic nucleus-nucleus collisions, bremsstrahlung of soft photons at forward rapidity traces the deceleration of incoming charges and is insensitive to the subsequent dynamical evolution. This makes it suited for constraining the initial conditions of the longitudinal net charge distribution. The idea of testing stopping via bremsstrahlung is as old as heavy-ion phenomenology~\cite{Kapusta:1977zb,Bjorken:1984sp,Dumitru:1993ph}. In the late 1990s, calculations of classical electromagnetic bremsstrahlung indicated that the expected effects are measurable in an experimentally accessible kinematic regime and that 
they could allow one to distinguish 
between different stopping scenarios~\cite{Jeon:1998tq,Kapusta:1999hb,Wong:2000hka}. 
This prompted studies for a dedicated forward detector at the BNL Relativistic Heavy Ion Collider (RHIC)~\cite{Jeon:1998tq} which, however, was not realized. As of today, forward bremsstrahlung from stopping of incoming charges remains a generally expected physics effect that has never been measured experimentally in heavy-ion collisions. 

One decade into the heavy-ion program at the CERN Large Hadron Collider (LHC), the ALICE Collaboration is currently investigating the physics opportunities of a next-generation TeV-scale heavy-ion detector that is based on ultrathin silicon technology~\cite{Adamova:2019vkf}. Besides high rate capabilities and excellent particle identification, this detector concept promises experimental access to observables at unprecedentedly low transverse momentum [$p_T \sim O(10\, {\rm MeV})$] and up to very forward rapidity  ($y=4$ or $y=5$), including prospects 
for soft and ultrasoft photon measurements. 
The present paper aims at initiating a discussion about the measurability of bremsstrahlung from stopping with this future heavy-ion collision experiment. To this end, we calculate the expected photon spectrum within the nominal acceptance of such a future detector
and we find that coverage in the range $10\, {\rm MeV/c} < p_T< 100\, {\rm MeV/c}$ should give access to a sizable yield.
While a full assessment of the experimental feasibility of such a bremsstrahlung measurement lies outside the scope of the present 
study and will depend on evolving detector studies, we shall find that the dominant photon background from meson decays has characteristically different distributions in transverse momentum and centrality. This should facilitate experimental strategies to isolate the effect. 

In the context of a next-generation heavy-ion experiment at the LHC~\cite{Adamova:2019vkf}, the prospects for soft and ultrasoft photon measurements have been discussed recently in the context of Low's theorem~\cite{Low:1958sn}. This theorem formally relates hadronic multiparticle production amplitudes without photons to expectation values for soft photon production by dressing all electrically charged in- and out-going lines of multiparticle production amplitudes with soft photon emissions.  Recent theoretical interest in these soft theorems arises from relating them to symmetries that reflect the infrared structure of gravity and gauge theory~\cite{Lysov:2014csa}. Low's theorem is a general quantum formulation of soft bremsstrahlung.  On general grounds, one expects that it interpolates as a function of resolution scale between the incoherent and the totally coherent limits of multiphoton bremsstrahlung. A classical formulation should apply for sufficiently long  wavelength when the entire system of charge $2\, Z$ acts coherently as a single emitter whose internal structure is not resolved by the emitted photons. Here, we work within the classical formalism of \cite{Jackson:1998nia} used previously in ~\cite{Jeon:1998tq,Kapusta:1999hb,Wong:2000hka}, and we check that this condition is met. The characteristic $1/p_T$ divergence
of photon bremsstrahlung is captured by this classical formulation, and our calculation addresses the question~\cite{Adamova:2019vkf} at which $p_T$-scale it will become experimentally accessible. 

\begin{figure*}
\begin{center}
  \includegraphics[width=0.98\textwidth]{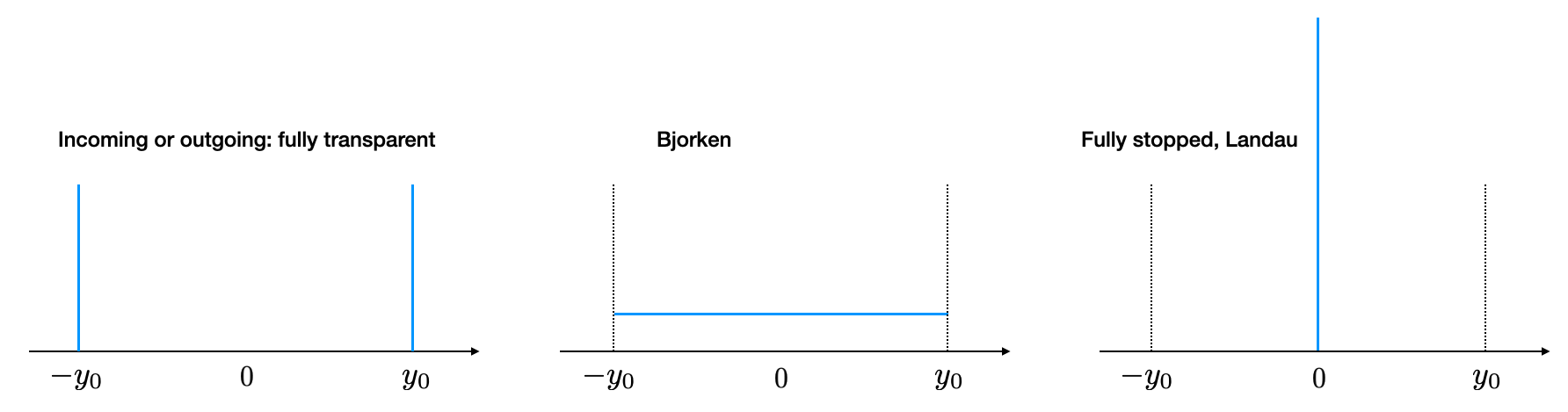}
\end{center}
\caption{Toy models of the final rapidity distribution of net charge in a nucleus-nucleus collision: 
i) if the collision were fully transparent, charges in the outgoing state would stay at incoming projectile rapidities $y_0$ and $-y_0$.
ii) in a Bjorken boost-invariant scenario, the charge distribution is flat iii) the case that charges are fully stopped in the center-of-mass
frame of the collision. 
} 
\label{fig1}
\end{figure*}
\section{Modelling the longitudinal charge distribution}
\label{sec2}
Incoming nuclear projectiles of charge $Z$ represent charge currents $J^{(in)}_{\pm}({\bf x},t)$ that propagate along the beam direction 
$z$ with  velocities $\pm v_0 = \pm \tanh y_0$ set by the beam rapidity $y_0$ which for all practical purposes can be identified with $\ln\left(\frac{\sqrt{s_{\rm NN}}}{m_N}\right)$,
\begin{equation}
    J^{(in)}_{\pm}({\bf x},t) = \pm Z\, e\, v_0\, \rho_{\rm in}(r)\, \delta(z \mp v_0t)\, \Theta(-t)\, .
    \label{eq1}
\end{equation} 
Here, ${\bf x} = ({\bf r},z)$ and we denote by $\rho_{\rm in}(r)$ the incoming charge density in the plane transverse to the beam. After the collision at time $t=0$, charges are partially stopped, \emph{i.e.}, they propagate with velocities $v(y)$ satisfying 
$-v_0 < v(y) < v_0$. The outgoing current takes the form
\begin{equation}
    J^{(out)}({\bf x},t) = \Theta(t)\, \int_{-y_0}^{y_0} \rho({\bf r},y,t)\, v(y)\, \delta\left(z - v(y)t \right)\, dy\, ,
    \label{eq2}
\end{equation} 
where  the charge density  $\rho({\bf r},y,t)$ is differential with respect to rapidity $y$. This density is normalized to the entire charge deposited in the collision region, $\int d{\bf r}\, \int_{-y_0}^{y_0} dy\, \rho({\bf r},y,t) = 2\, Z\, e$.
After the start of the collision at $t=0$, charges will be decelerated rapidly and they will be stopped in their final rapidity window after a very short time $t_s$, \emph{i.e.},  $\rho_{\rm out}({\bf r},y) = \rho({\bf r},y,t)\vert_{t>t_s}$. 
We work in natural units ($c=\hbar=1$).

The intensity and number of photons of energy $\omega$ radiated into the angular opening $d\Omega$ can be obtained from the classical bremsstrahlung formula~\cite{Jackson:1998nia} 
\begin{equation}
	\frac{d^{2} I}{d \omega d \Omega}=\omega \frac{d^{2} N}{d \omega d \Omega}=|\mathbf{A}|^{2}\, ,
	\label{eq3}
\end{equation}
where $\mathbf{A}$ is defined in terms of the current $\mathbf{J}(\mathbf{x}, t)$ and the direction ${\bf n}$ of the outgoing photon
\begin{equation}
\mathbf{A}(\mathbf{n}, \omega)= -\frac{i\omega}{2\pi}
\int dt\int d^{3}x\,  \mathbf{n} \times(\mathbf{n} \times \mathbf{J}(\mathbf{x}, t)) e^{i \omega(t-\mathbf{n} \cdot \mathbf{x})}\, .
\label{eq4}
\end{equation}
The current sums over all in- and outgoing contributions 
\begin{equation}
\mathbf{J} = \mathbf{J}^{(in)}_{+} + \mathbf{J}^{(in)}_{-} + \mathbf{J}^{(out)}\, .
\end{equation}
For the problem under consideration, 
$\mathbf{J}$ points always along the beam direction $\vec{e}_z $ and the direction of 
$\mathbf{n}=\vec{e}_\varphi \sin \theta  + \vec{e}_z \cos \theta $  with $\vec{e}_\varphi  \perp \vec{e}_z $ lying in the 
transverse plane.  For the discussion of experimental acceptances, it is useful to convert into
 pseudorapidity  
 \begin{equation}
 \eta = - \ln \left[ \tan\left(\tfrac{\theta}{2}\right)\right]\, .
 \end{equation}
The current entering \eqref{eq4} depends on the outgoing charge distributions $\rho({\bf r},y,t)$. It is in this way that
classical bremsstrahlung becomes a tool for constraining initial longitudinal conditions. 

We consider first the simple scenarios depicted in Fig.~\ref{fig1}:
\begin{enumerate}
\item Full transparency: charges are not decelerated, 
\begin{equation}
\frac{d^{2} I}{d \omega d \Omega} =0\, .
\label{eq5}Eq.
\end{equation}
\item Bjorken-stopping: 
\begin{equation}
\rho({\bf r},y,t) = \frac{Z\, e }{y_0} \, \rho_{\rm in}({\bf r})\, \Theta(t)\, \Theta(y_0-|y|)\, .
\label{eq6}
\end{equation}
\item Landau-stopping: 
\begin{equation}
\rho({\bf r},y,t) = 2\, Z\, e\,  \rho_{\rm in}({\bf r})\, \Theta(t)\, \delta(y)\, .
\label{eq7}
\end{equation}
\end{enumerate}
To appreciate the usefulness of such simple models, let us consider briefly the \emph{hypothetical scenario} of a 
bell-shaped charge rapidity distribution  $\rho_{\rm out}(y)$ which would amount to less (more) charge deceleration than 
Eq.~\eqref{eq7} [Eq.~\eqref{eq6}], respectively. One therefore expects
that the bremsstrahlung spectrum of this bell-shaped distribution is bracketed by the cases of Eqs.~\eqref{eq6} and \eqref{eq7}. 
A minimal prerequisite for being sensitive to bremsstrahlung from stopping is then that the scenario of \eqref{eq7} can be 
distinguished  from the null-hypothesis \eqref{eq5}, and a measurement that can disentangle 
the scenarios ~\eqref{eq6} and \eqref{eq7} demonstrates sensitivity to distinguish between different conceivable stopping scenarios.

In the simple scenarios of Eqs.~\eqref{eq6} and \eqref{eq7}, charges are assumed to be stopped instantaneously and stopping is independent of radial position. In general, charges will decelerate over a finite time $\Delta t_f$ and they may decelerate differently
at different radial positions. There is, however, a simple parametric reason for why these details should have a negligible effect on
bremsstrahlung radiation: Pb ions at LHC have a gamma-factor $\gamma \approx 2700$ which makes them Lorentz-contracted 
pancakes of longitudinal thickness $\approx 0.005$ fm in the rest-frame of the collision. Any stopping must be completed before the charges have traversed the other nucleus, \emph{i.e.}, it must be completed within a time $\approx 0.005$ fm/c. To be sensitive to
the detailed time- and/or position dependence of stopping, \emph{forward} bremsstrahlung photons would have to resolve this Lorentz-contracted thickness. However, this is not possible with the photon energies $\omega $ that we consider in the following and for which 
$1/\omega \gg 0.005$ fm/c. Consistent with this simple parametric argument, we have found for $\omega < 2$ GeV only very small 
($< 5 \%$) numerical differences between the sophisticated position- and time-dependent stopping scenario with Bjorken 
boost-invariant final charge distribution 
considered in Ref.~\cite{Kapusta:1999hb}, and the simplified model ~\eqref{eq6} considered here (data not shown). 

We note that this argument applies only to photons at sufficiently forward rapidity that would need to resolve the strongly Lorentz-contracted \emph{longitudinal} structure of the nucleus. In contrast, if emitted at central rapidity (e.g., emitted close to mid-rapidity 
$\theta = 90^\circ$), a $\omega = 200$ MeV photon resolves $O(1)$ fm distances in the \emph{transverse} direction in which the 
nucleus is not Lorentz-contracted. Therefore, photon emission around central mid-rapidity is expected to be sensitive to the
internal structure of the charge distribution, while soft photon emission at forward rapidity is expected to be described by the 
classical formulation recalled here. 

\section{Numerical results}
In the last section, we gave qualitative arguments for how the classical bremsstrahlung \eqref{eq3} from the incoming 
and outgoing charge currents  \eqref{eq1}, \eqref{eq2} provides insight into the longitudinal rapidity 
distribution of net charges in the \emph{initial stage} of the collision. In this section, we discuss the corresponding spectra
and  we provide numerical results. 

Both the Bjorken and the Landau stopping scenarios in Eqs.~\eqref{eq6} and \eqref{eq7} lead to charge distributions for
which the dependence on transverse radius and rapidity factorizes, $\rho_{\rm out}({\bf r},y) = \rho_{\rm in}({\bf r})\, \rho(y)$. 
Inserting this ansatz into \eqref{eq2}, one finds
\begin{eqnarray}
\frac{d^2I}{d\omega d\Omega} &=& \frac{\alpha Z^2}{4\pi^2}\sin^2\theta \left| F(\omega \sin\theta) \right|^2 
\nonumber \\ 
&&\hspace{-.2cm}\times \left|
\left[ \int dy \frac{v(y) \rho(y)}{1-v(y)\cos\theta} - \frac{2v_0^2 \cos\theta}{1-v_0^2
\cos^2\theta} \right] \right| ^2 ,
\label{eq8}
\end{eqnarray}
where $v(y) = \tanh(y)$ and where $F$ denotes the transverse nuclear form factor, 
\begin{equation}
F(\omega \sin\theta)=
\int d^2r_{\perp} \, \rho_{\rm in}\left( r_{\perp} \right)
e^{-i \omega {\bf n}\cdot {\bf r}_{\perp}}\, .
\label{eq9}
\end{equation}
In the product $\rho_{\rm out}({\bf r},y) = \rho_{\rm in}({\bf r})\, \rho(y)$, we take the longitudinal rapidity distribution normalized to 2,
so that the scenarios discussed in Sec. \ref{sec2} and sketched in Fig.\ref{fig1} correspond to 
\begin{equation}
	\rho(y) =   
	\left\{ \begin{array}{cc} \delta(y-y_0) +  \delta(y+y_0)  &
	 \hbox{(transparent)}\\ \tfrac{1}{y_0} \Theta(y_0-|y|) & \hbox{(Bjorken)} \\ 2\delta(y) & \hbox{(Landau)} 
	 \\ \tfrac{2}{\sigma \sqrt{2\pi}} \exp\left[ - \tfrac{y^2}{2\sigma^2}  \right]  & \hbox{(Gaussian)}\end{array} \right.
	 \label{eq10}
\end{equation}
For the simplifying assumption that charges in heavy nuclei are distributed homogenously in a sphere of radius $R$,
the form factor depends only on the dimensionless scale $q\equiv\omega R \sin\theta$~\cite{Kapusta:1999hb} 
\begin{eqnarray}
F(q) &=& \frac{3}{q^2}\left( \frac{\sin q}{q} - \cos q \right)    \quad \hbox{[fixed sphere]} \, ,
\label{eq11}\\
&& q\equiv\omega R \sin\theta\, .
\nonumber
\end{eqnarray}
Equation~\eqref{eq9} could be evaluated easily for more refined distributions, such as e.g. a Woods-Saxon-distribution.
However, as argued above, mild differences in transverse profiles should not affect longitudinal bremsstrahlung spectra 
significantly and we therefore prefer to work with the simple analytic formula \eqref{eq11}. 
%
\begin{figure}[h] 
\begin{center}
\includegraphics[width=0.33\textwidth]{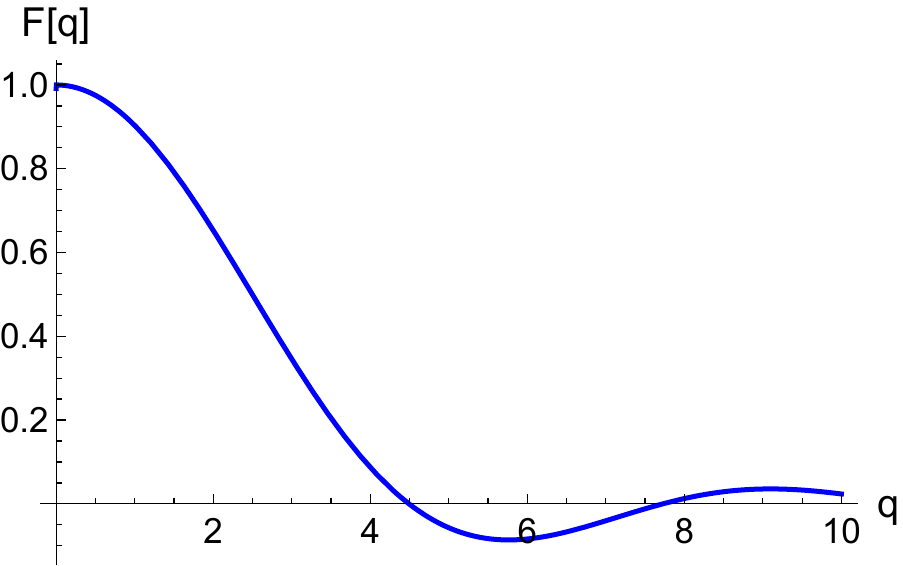}
\end{center}
\caption{The form factor \eqref{eq11} of the transverse nuclear charge distribution as a function of $q\equiv\omega R \sin\theta$.} 
\label{fig2}
\end{figure}
%

The form factors \eqref{eq9} satisfy by construction $\lim_{q \rightarrow 0}\, F(q) = 1$. For the case of a fixed sphere, Eq.~\eqref{eq11}, the corresponding form factor $F(q)$ is plotted in Fig.~\ref{fig2}. The characteristic fall-off properties of $F(q)$
(such as  $F(q) \geq 0.9$ for $q< 1$ and $F(q) < 0.1$ for $q>4$) can be expected to hold for a broad class of realistic charge distributions. 
In the following, we are interested in forward bremsstrahlung. For a Pb nucleus with $R = 6.8$ fm and for pseudorapidity $\eta =3$ ($\eta =5$), the condition $q<1$ translates into $\omega < 1/(R\sin\theta) = 1.5 \tfrac{1}{\rm fm}  \approx 300\, {\rm MeV}$ ($\omega < 2.2$ GeV), respectively. In this kinematic regime, on which our discussion will focus, the squared form factor in \eqref{eq8} corresponds therefore to a mild (20 \% or less) deviation from unity. 

%
\begin{figure*}[htbp] 
\begin{center}
  \begin{tabular}{ccc} 
\includegraphics[width=0.33\textwidth]{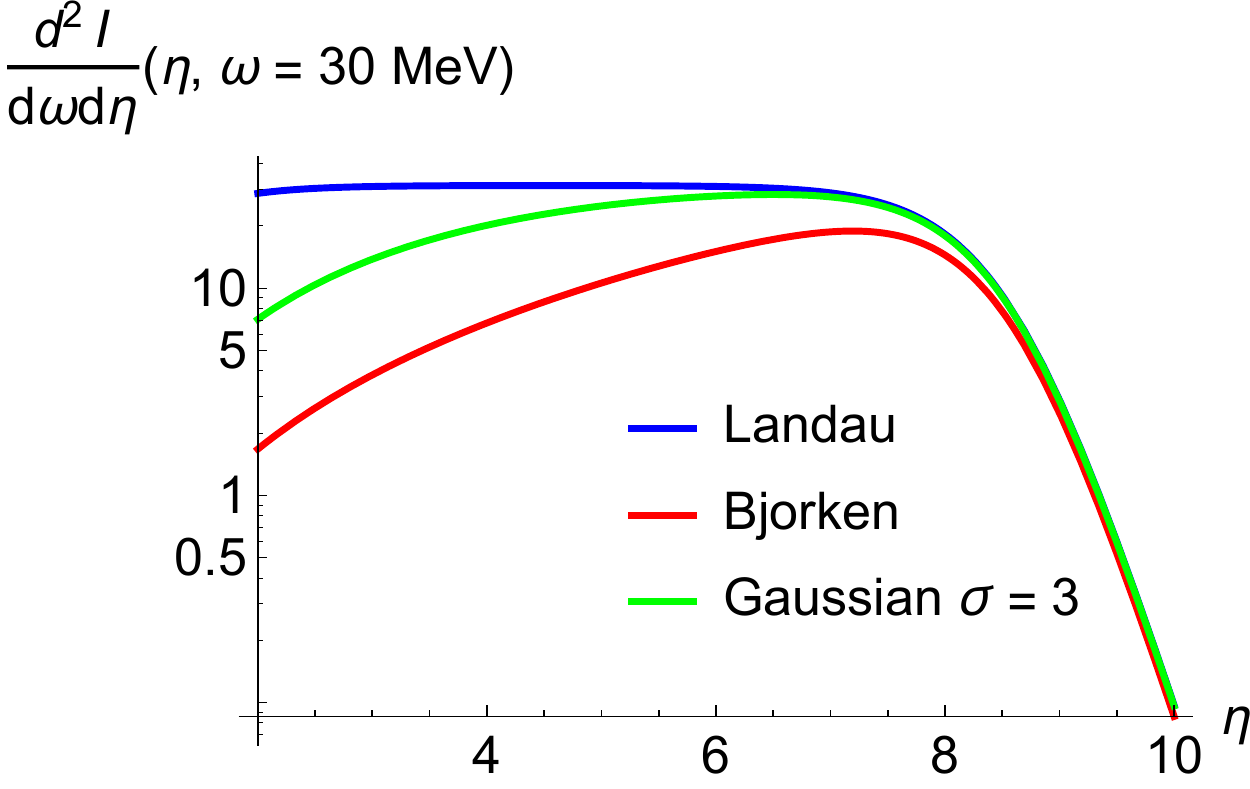}
  &
\includegraphics[width=0.33\textwidth]{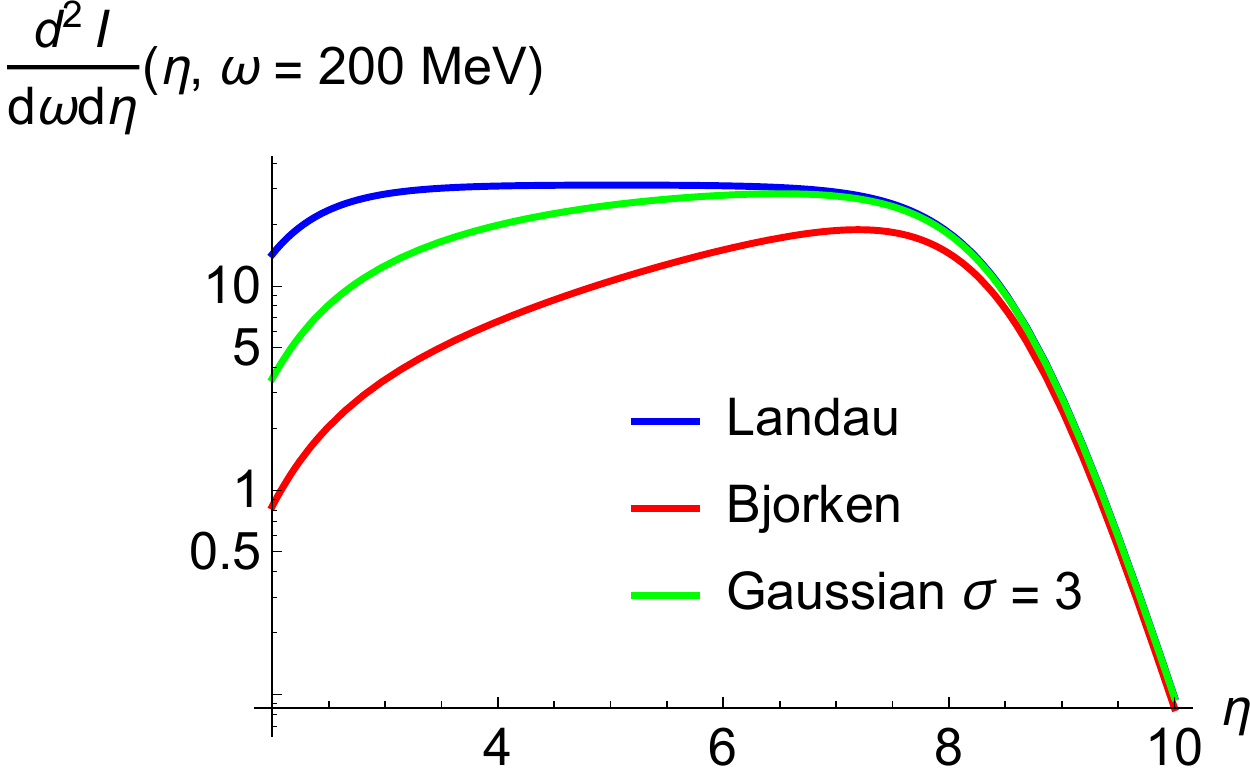}
  &
\includegraphics[width=0.33\textwidth]{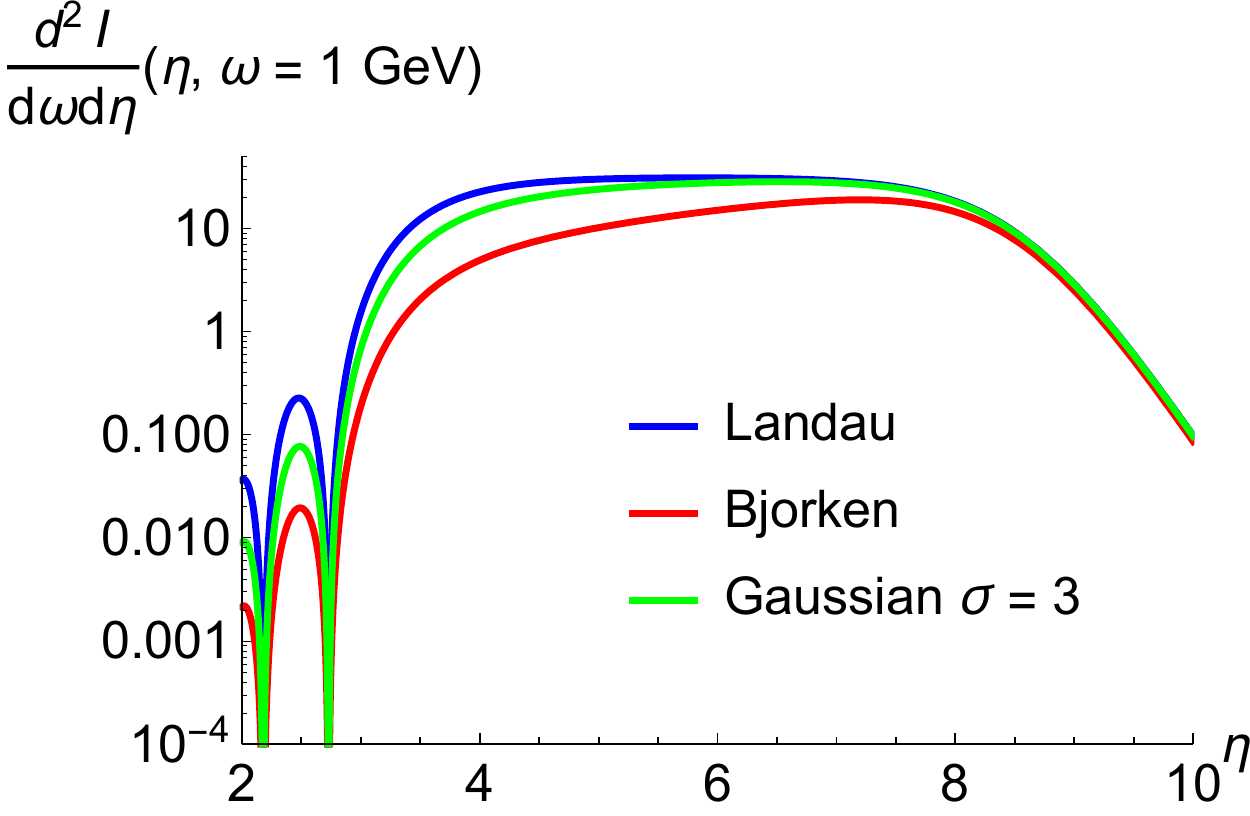}
  \end{tabular}
\end{center}
\caption{The double-differential photon energy distribution $\tfrac{d^2I}{d\omega\, d\eta}$ for different photon energies $\omega$
as a function of pseudorapidity $\eta$. Results are shown for the three different stopping scenarios ~\eqref{eq10}
in central PbPb collision at $\sqrt{s_{\rm NN}}= 5.02$ TeV. } 
\label{fig3}
\end{figure*}
%
\begin{figure*}[htbp] 
\begin{center}
  \begin{tabular}{ccc} 
\includegraphics[width=0.33\textwidth]{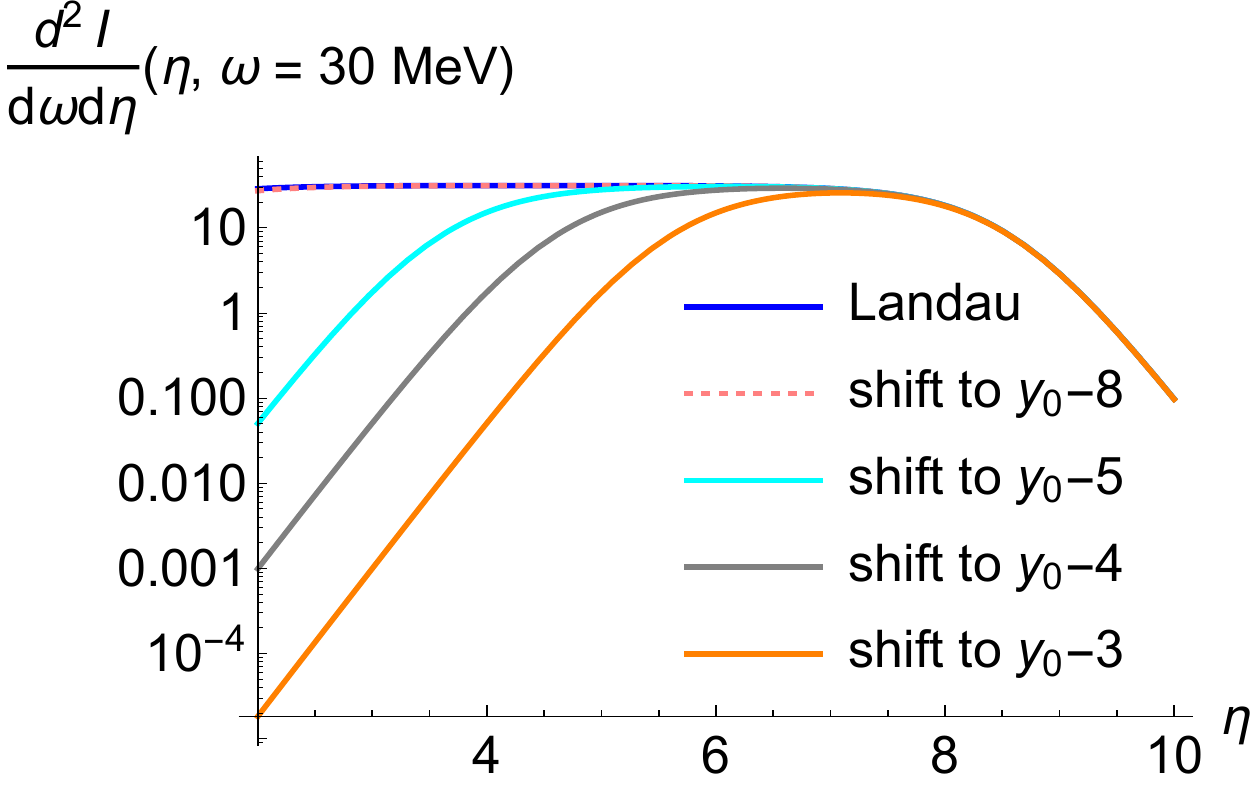}
  &
\includegraphics[width=0.33\textwidth]{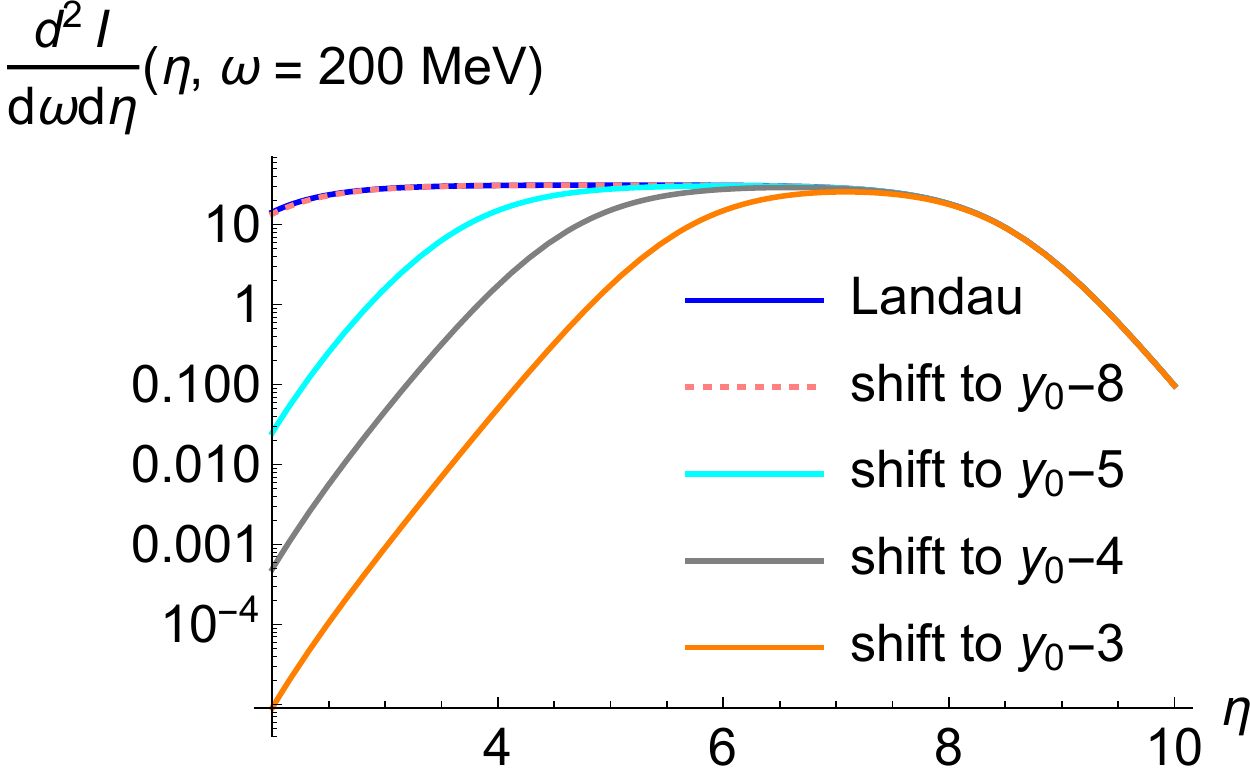}
  &
\includegraphics[width=0.33\textwidth]{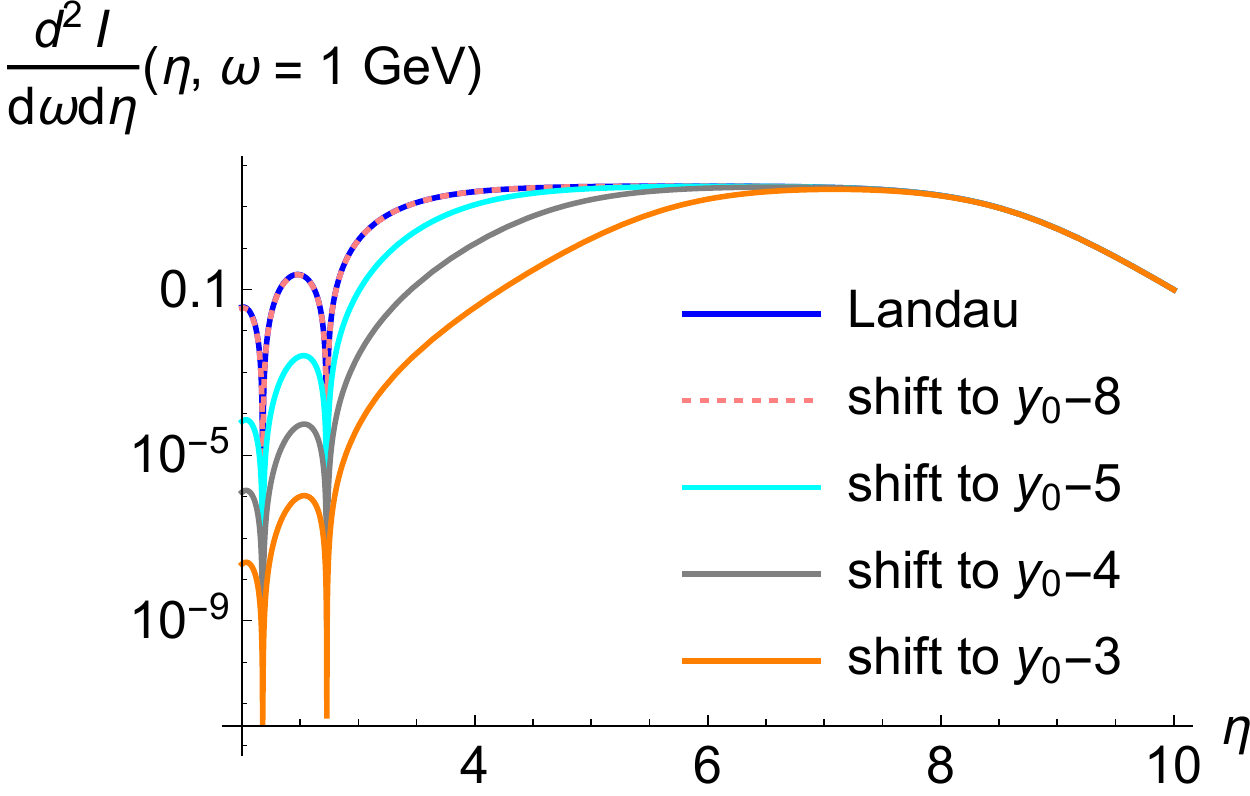}
  \end{tabular}
\end{center}
\caption{Same as Fig.~\ref{fig3}, but for the stopping scenarios of Eq.~\eqref{eq16}.} 
\label{fig4}
\end{figure*}
\subsection{The pseudorapidity distribution}

We start by discussing the photon energy distribution 
\be
\frac{d^2I}{d\omega d\eta} = \frac{d^2I}{d\omega \sin\theta d\theta d\phi}\, 2\pi\,   \sin^2\theta
\label{eq13}
\ee
which we obtain from Eq.~\eqref{eq8} by integrating over the
azimuth $d\phi$, using 
\be
\frac{d\theta}{d\eta} 
= \frac{1}{\cosh \eta} =
\sin\theta\, .
\ee

Figure~\ref{fig3} plots Eq.~\eqref{eq13} for different values of photon energy $\omega$ as a function of pseudorapidity. We first discuss 
the kinematic region of very forward pseudorapidity ($\eta > 8$, say), before turning to the features seen at smaller $\eta$. 

In $\sqrt{s_{\rm NN}} = 5.02$ TeV Pb-Pb collisions, the projectile rapidity is $y_0 = \ln\left(\frac{\sqrt{s_{\rm NN}}}{m_N}\right) = 8.5$. 
As seen from Fig.~\ref{fig3}, the bremsstrahlung energy distribution $\frac{d^2I}{d\omega d\eta}$ extends unattenuated up to comparable values of pseudorapidity $\eta$.  However, pseudorapidity  $\eta = - \ln \left[ \tan\left(\tfrac{\theta}{2}\right)\right]$ measures a polar angle, and a small  amount of photon bremsstrahlung is emitted at any arbitrarily small forward angle $\theta$. This is the reason for why there is 
energy at $\eta > 8.5$, though the energy decreases rather sharply with increasing $\eta$. 

At relatively high photon energy $\omega = 1$ GeV and relatively low pseudorapidity, Fig.~\eqref{fig3} displays two peculiar 
dips in the energy distribution around $\eta = 2.2$ and $\eta = 2.8$. As we discuss now, these are artifacts of our simple fixed sphere
model \eqref{eq11} for the charge distribution, and these artifacts may help to illustrate the range of validity of our calculation.
To clarify this point, we recall our comments about Fig.~\ref{fig2}: the simple form factor $F(q)$ used in our calculation is expected to 
have a very small model-dependence for $q < 1$, but it will be completely model-dependent for $q >4$ where it shows peculiar zero-crossings. Indeed, for $\omega = 1$ GeV, $R=6.8$ fm the Pb radius used in our calculation and $q=4$, we find $\sin\theta = \tfrac{q}{\omega R} = \tfrac{4}{5\, \times 6.8} = 0.12$ which corresponds exactly to $\eta = 2.8$. The dips seen in Fig.~\ref{fig3} for $\omega = 1$ GeV are in one-to-one correspondence with the zero-crossings of $F(q)$ for $q>4$. It also follows from $\sin\theta = \tfrac{q}{\omega R}$
that for $\omega = 1$ GeV, values $q < 1$ correspond to $\eta > 4.2$. This is the region in which we expect our calculation to yield physical results.  

For softer photon energies, $\omega = 200$ MeV say, the same argument implies that a dip should show up at $\eta = 1.1$. This dip exists but it  is not displayed in Fig.~\ref{fig3}, since we plot only for $\eta > 2$. At this lower photon energy, $q < 1$ corresponds to $\eta > 2.6$. In general, the softer the photon energy, the less sensitive is our calculation to geometrical details and the more it can be trusted over a wide range of pseudorapidity.

\begin{figure*}[htbp] 
\begin{center}
  \begin{tabular}{ccc} 
	\includegraphics[width=0.33\textwidth]{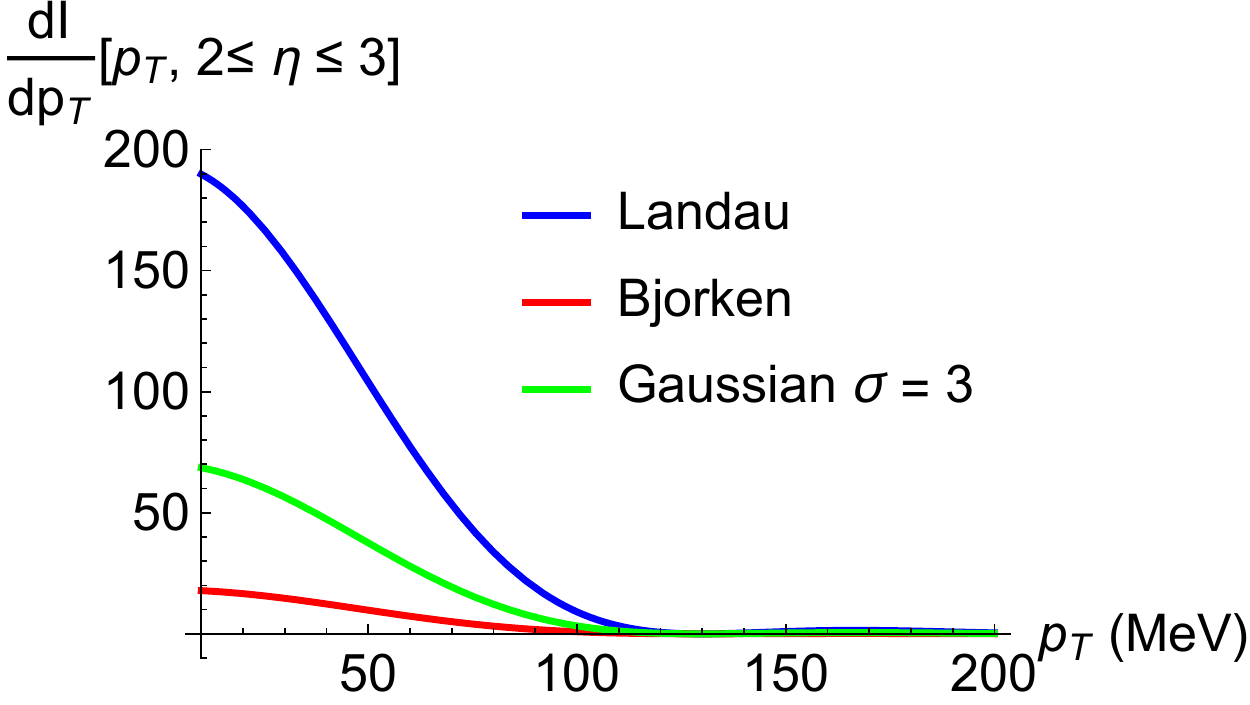}
  &
  \includegraphics[width=0.33\textwidth]{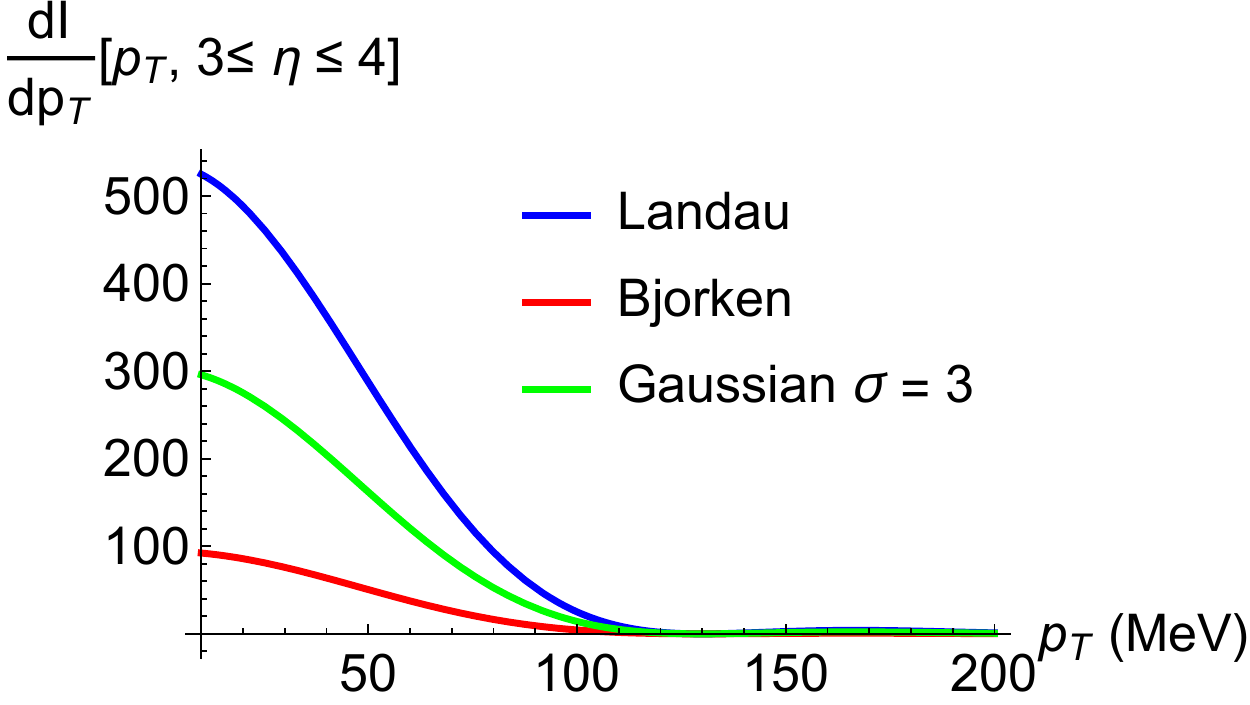}  
    &
  \includegraphics[width=0.33\textwidth]{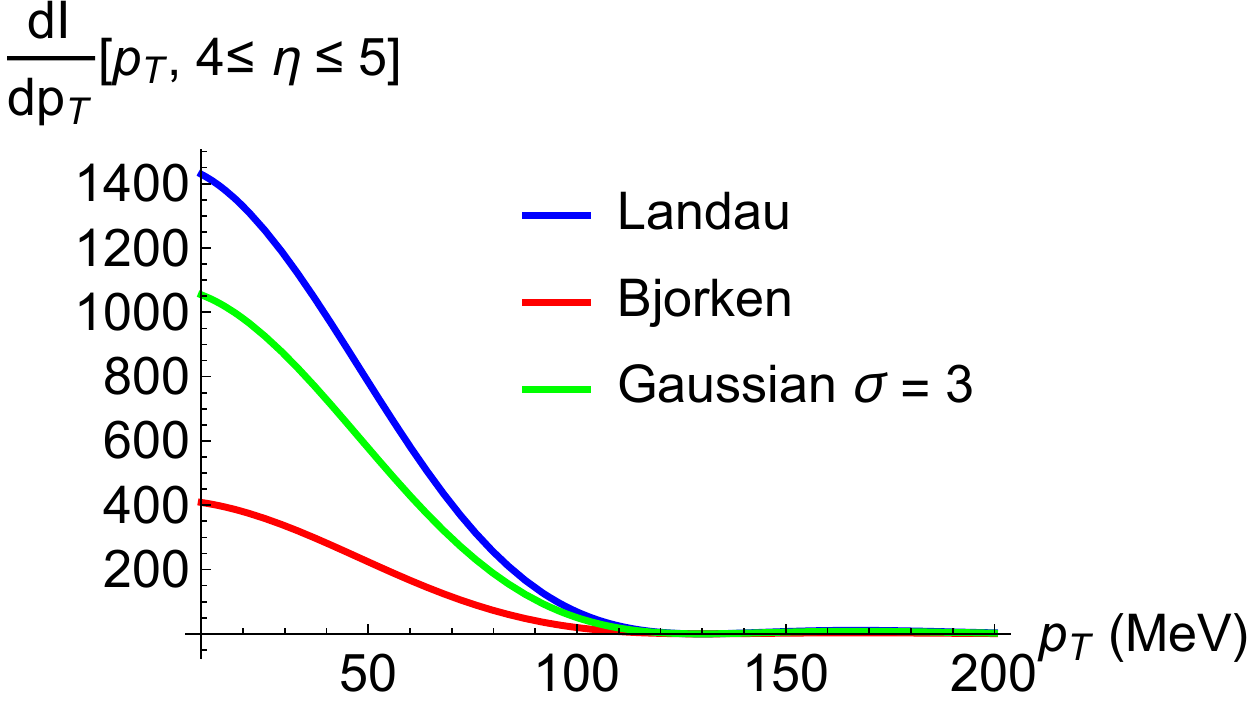}  
  \end{tabular}
\end{center}
\caption{The $p_T$-dependence of the photon bremsstrahlung distribution $\int_{\eta_-}^{\eta_+} \frac{d^2I}{dp_T d\eta} d\eta$, integrated over different windows $\left[\eta_-,\eta_+ \right]$ of pseudorapidity. The shape of the  $p_T$-distribution remains almost unchanged, but the yield
increases steeply towards forward rapidity.  
 } 
\label{fig5}
\end{figure*}

Within the range of forward pseudorapidity in which our calculation is expected to be model-independent ($\eta > 3$ or $\eta > 4$, depending on photon energy), the three models of charge stopping displayed in Fig.~\ref{fig3} lead to energy distributions that are
numerically different and that differ in their rapidity dependence. To illustrate the physics behind these differences, it is useful to 
introduce another class of stopping scenarios, in which all net charges are shifted by the same fixed number of units $y_{\rm shift}$ in rapidity,
\be
	\rho(y) = \delta\left(y-(y_0-y_{\rm shift})\right) + \delta\left(y+(y_0-y_{\rm shift})\right)\, .
	\label{eq16}
\ee
From the resulting energy distributions in Fig.~\ref{fig4}, we conclude that for a stopping scenario \eqref{eq16}, the soft photon energy distribution at fixed $\omega$ forms a plateau within the pseudorapidity range $y_0-y_{\rm shift} \lesssim \eta \lesssim y_0$. 
The more the net charge is stopped, the more the radiation extends towards mid-rapidity. Any model of longitudinal stopping that is described by a continuous function $\rho(y)$ may be viewed as a linear superposition of distributions \eqref{eq16}. This explains why for sufficiently soft photons, the energy distributions in Fig.~\ref{fig3} is flat for the Landau case, but rises with increasing $\eta$ for 
models with continuous final longitudinal charge distribution $\rho(y)$. In this sense, the $\eta$-distribution of bremsstrahlung photons monitors the rapidity-dependence of stopped charges. 

So far, we have discussed  bremsstrahlung in terms of a double-differential distribution in $\omega$ and $\eta$. To discuss issues 
of experimental acceptance and measurability, it is preferable to switch to $p_T$ and $\eta$, 
\be
\frac{d^2I}{dp_T d\eta} = \frac{d^2I}{d\omega d\Omega}\, 2\pi\,   \sin^2\theta\, \cosh\eta\, .
\ee
Figure~\ref{fig5} shows the corresponding $p_T$-differential photon energy distribution integrated over $\eta \in \left[\eta_-;\eta_+ \right]$, 
$ \int_{\eta_-}^{\eta_+} \frac{d^2I}{dp_T d\eta} d\eta$. This plot makes it clear that experimental access to bremsstrahlung photons requires acceptance for $p_T <100$ MeV. This is so irrespective of pseudorapidity.

Integrating the spectra in Fig.~\ref{fig5} over $p_T$, we find that for the Landau stopping 
scenario, a total of 64 GeV (23.5 GeV, 8.5 GeV) energy is radiated per central Pb-Pb collision into the phase space region 
$10\, {\rm MeV} < p_T < 100\, {\rm MeV}$ and $\eta \in \left[4,5\right]$   ($\left[3,4\right]$, $\left[2,3\right]$), respectively. For the other stopping scenarios plotted in Fig.~\ref{fig5}, the total energy radiated into these three phase space regions is accordingly smaller
(47, 13 and 3 GeV for the Gaussian scenario, and 18, 4 and 0.8 GeV for the Bjorken scenario).

\subsection{The photon number distribution}

\begin{figure*}[htbp] 
\begin{center}
  \begin{tabular}{ccc} 
	\includegraphics[width=0.33\textwidth]{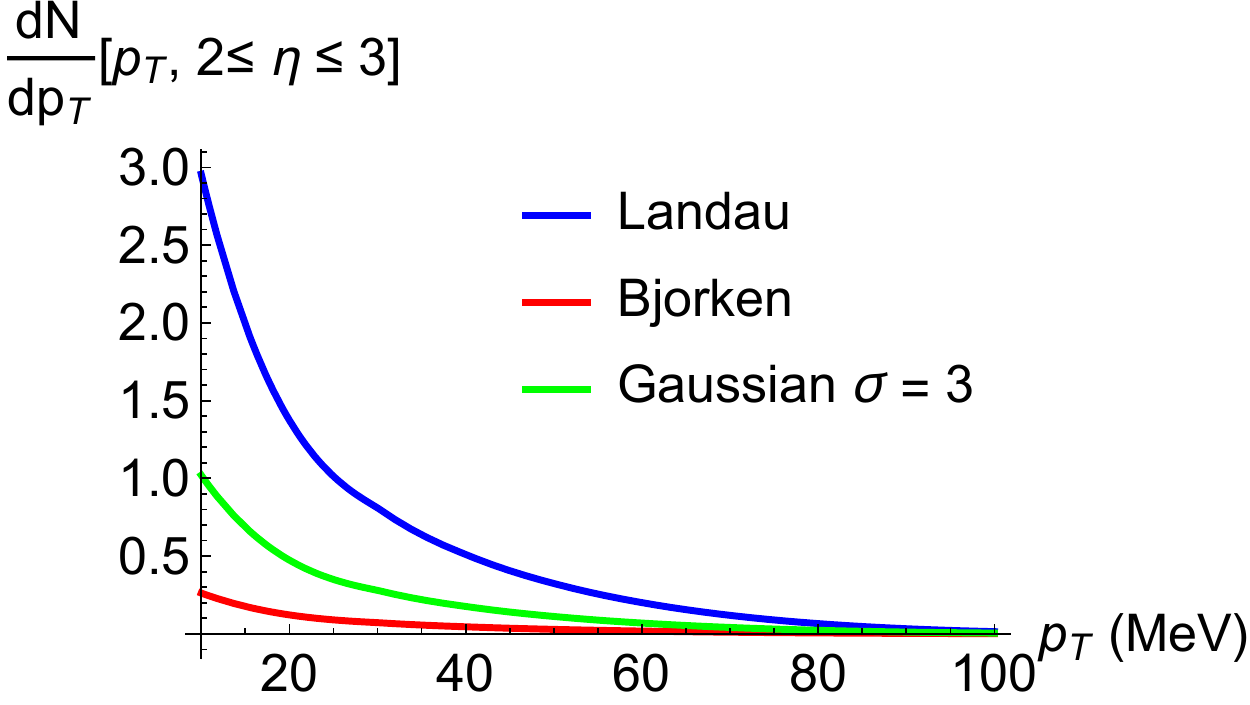}
  &
  \includegraphics[width=0.33\textwidth]{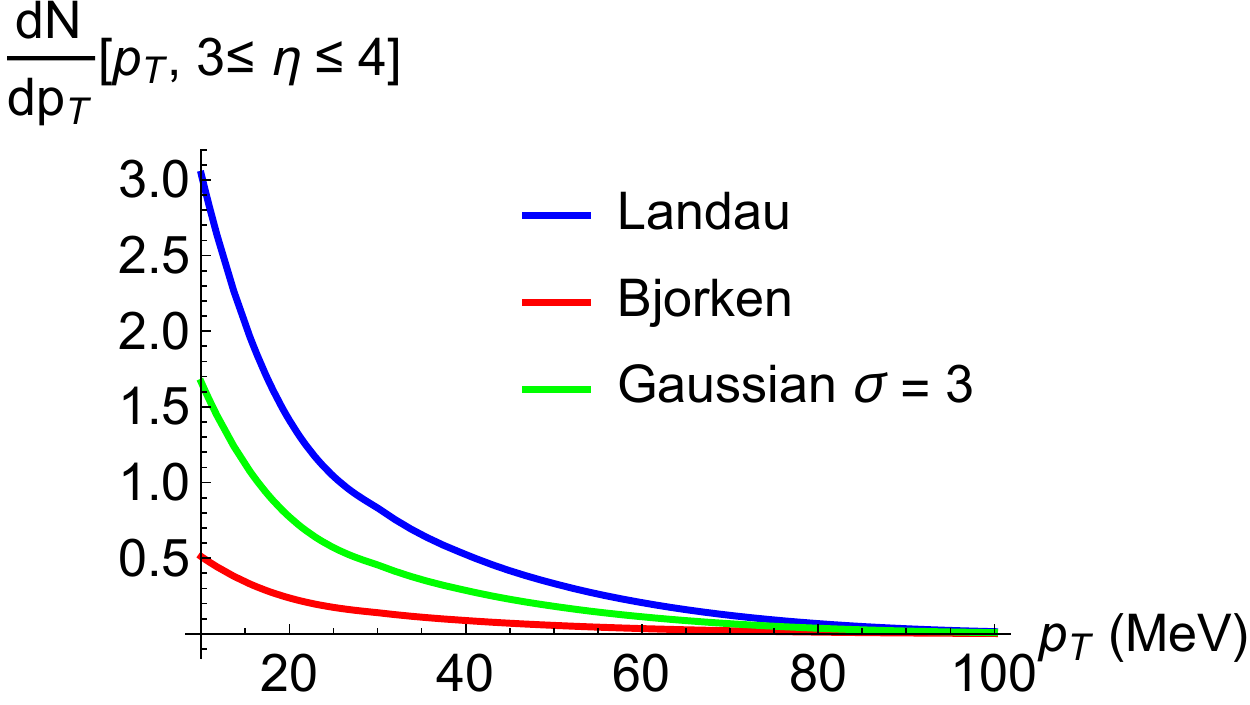}  
    &
  \includegraphics[width=0.33\textwidth]{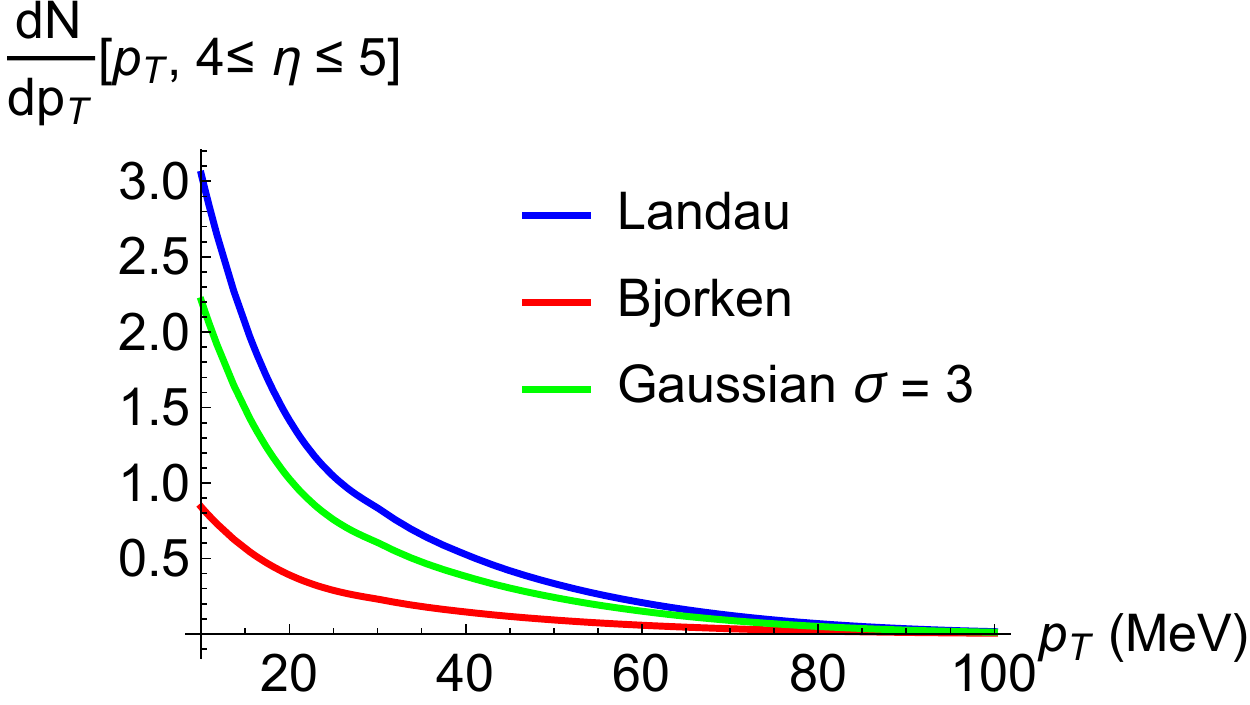}  
  \end{tabular}
 \begin{tabular}{ccc}
\includegraphics[width=0.33\textwidth]{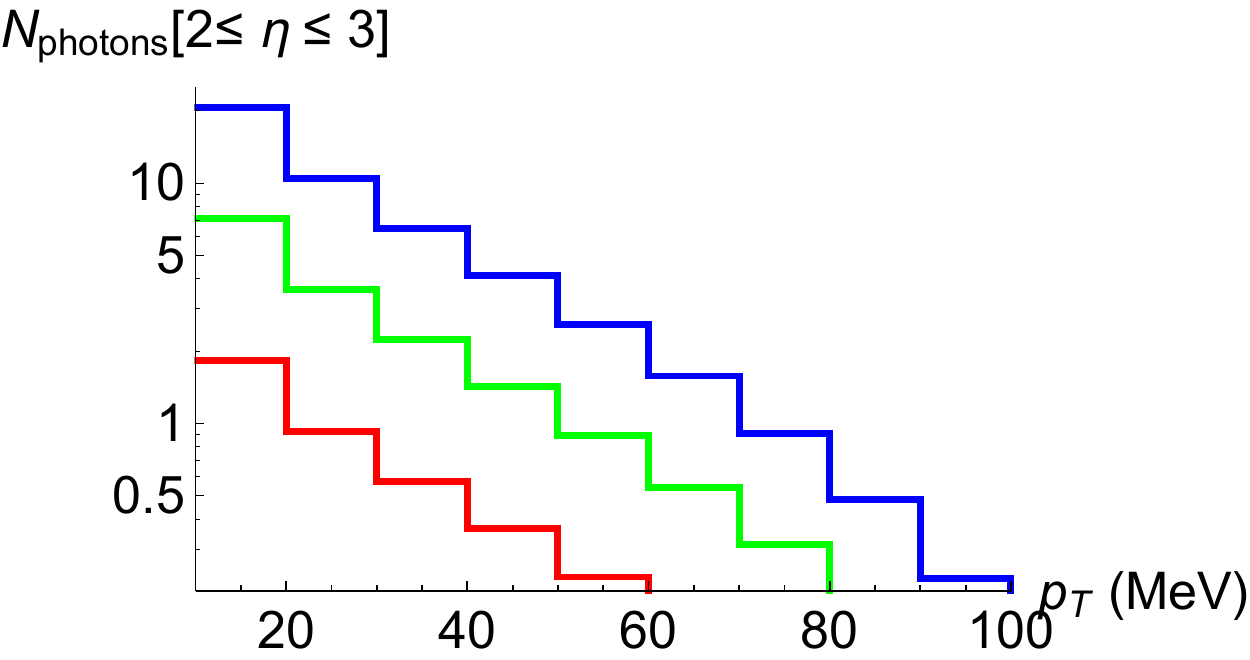}
  &
  \includegraphics[width=0.33\textwidth]{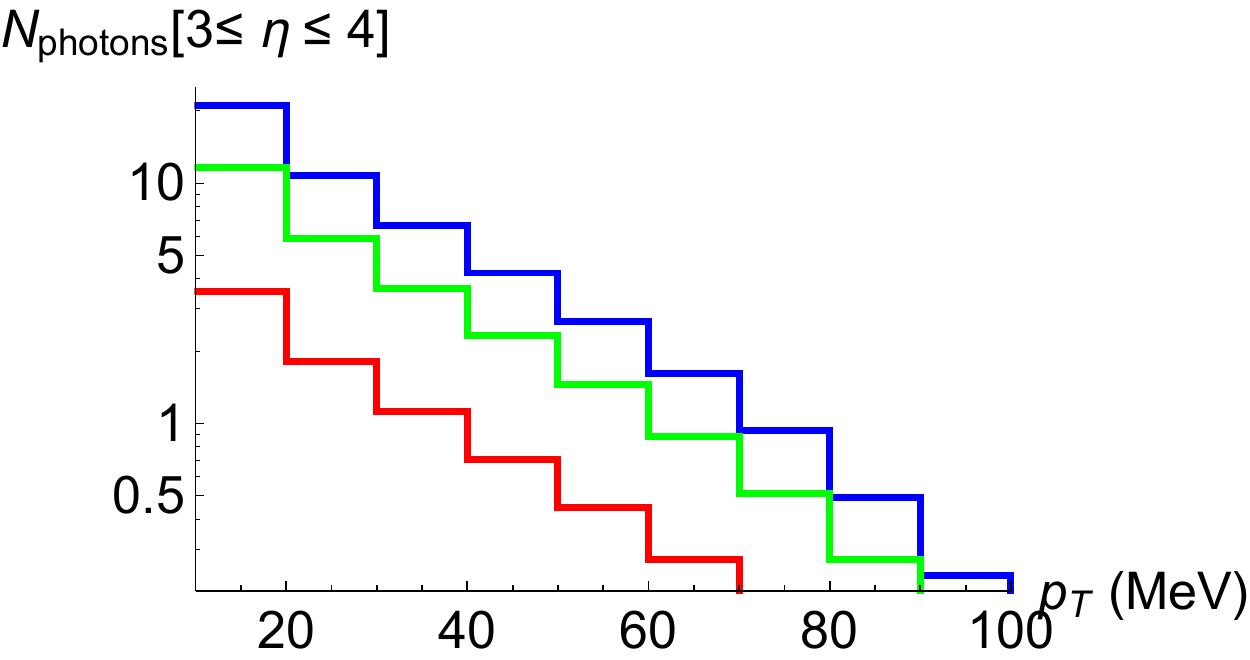}  
    &
  \includegraphics[width=0.33\textwidth]{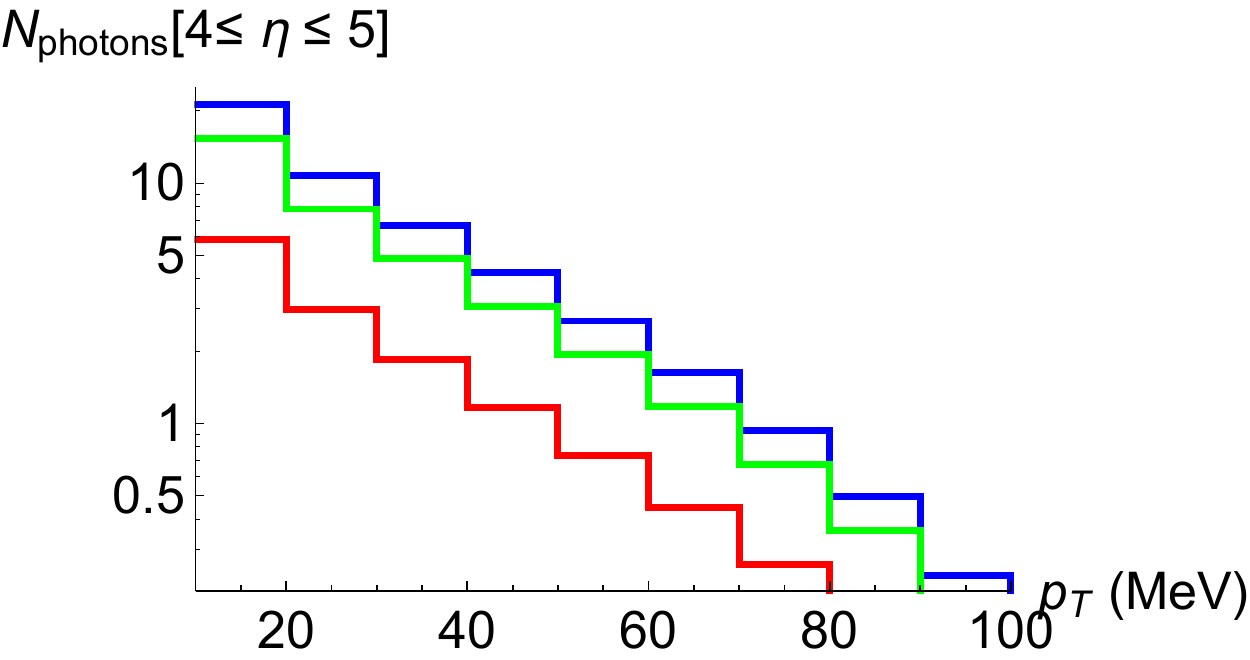}  
  \end{tabular}

\end{center}
\caption{Upper panel: the differential photon number spectrum~\eqref{eq18} as a function of transverse momentum for different pseudorapidity bins. Lower panel: the corresponding number of photons in bins of 10 MeV in transverse momentum.
 } 
\label{fig6}
\end{figure*}

We finally translate the results shown above into the number of photons radiated per unit phase space,
\be
	\frac{d^2N}{dp_T d\eta} = \frac{1}{p_T\, \cosh\eta} \frac{d^2I}{dp_T d\eta} \, .
	\label{eq18}
\ee
Depending on the stopping scenario, we find between 5 and 20 photons per unit pseudorapidity in the range 
$p_T \in \left[10\, {\rm MeV}; 20\, {\rm MeV}\right]$, see Fig.~\ref{fig6}. With increasing $p_T$, the number of
bremsstrahlung photons decreases, and in the $p_T$-bin $\left[50\, {\rm MeV}; 60\, {\rm MeV}\right]$, we find
between 0.5 and 5 bremsstrahlungs photons per unit pseudorapidity and per event. 

We find that for $p_T \lesssim 40$ MeV, the photon number spectrum  in Fig.~\ref{fig6} follows the characteristic 
$dN/dp_T \propto 1/p_T$ dependence of soft photon radiation while it decays somewhat more steeply at higher $p_T$.

\subsection{``Background" photons}
The question of whether and how bremsstrahlung photons can be disentangled from other sources of photons requires 
detailed event generator studies that lie outside the scope of the present exploratory calculations. Here, we restrict ourselves
to some qualitative considerations: 

Photons from $\pi^0$-decays  are expected to provide the most important background. Also $\eta$-mesons have a 
branching ratio of $40\%$ into two photons and need to be included in a realistic cocktail.\footnote{In addition to resonance decays, forward photons are also radiated off electron-positron pairs produced in the colliding Weizs\"acker-Williams fields~\cite{Hencken:1999xw}. The centrality dependence of this contribution is different from that of Eq.~\eqref{eq8}.}
 It is known that in PbPb
collisions at the LHC, these mesons have the same nuclear modification factor as charged 
pions in the range $1\, {\rm GeV} < p_T < 20\, {\rm GeV}$ and around mid-rapidity~\cite{ALICE:2018mdl}. For the following
simple estimates, we therefore assume that the $\pi^0$- and the $\pi^\pm$-distributions are the same for all rapidities 
and for all transverse momenta. 

At mid-rapidity, the $p_T$-differential charged pion spectrum $dN^{\pi^\pm}/dp_T\, dy$ is approximately constant,
$dN^{\pi^\pm}/dp_T\, dy \simeq 2000/{\rm GeV}$ for $p_T < 500$ MeV at $y=0$ (see e.g. Fig.21 of Ref.~\cite{Nijs:2020roc}
which replots data from ~\cite{ALICE:2014juv}).
In central PbPb collisions at the LHC, $dN_{\rm ch}/d\eta$  decreases by almost a factor two from 
$\eta =0$ to $\eta = 5$~\cite{ALICE:2013jfw}. We therefore expect
$dN^{\pi^0}/dp_T\, dy \simeq 500/{\rm GeV}$ for $p_T < 500$ MeV and $4 < \eta < 5$. This amounts to an approximately flat, 
$p_T$-independent distribution of five $\pi^0$s per event and per 10 MeV-bin, to be compared to a steeply falling 
$p_T$-distribution of a comparable number of bremsstrahlung photons in the range $p_T< 100$ MeV, see Fig.~\ref{fig6}. 
Our simple considerations thus indicate 
two characteristic differences between bremsstrahlung photons and ``background photons"
\begin{enumerate}
	\item \emph{characteristically different $p_T$-dependence}\\
	The $dN/dp_T$-spectrum of bremsstrahlung photons falls off $\propto 1/p_T$ or steeper in the range $p_T < 100$ MeV. 
	In contrast, light mesons that decay into photons have an approximately flat, $p_T$-independent $dN/dp_T$ distribution 
	in a wider $p_T$-range (up to $p_T < 500$ MeV, say). Bremsstrahlung photons should thus be visible as a characteristic 
	low-$p_T$ enhancement above a smooth almost $p_T$-independent baseline.
	\item  \emph{characteristically different centrality-dependence}\\
	The yield of bremsstrahlung photons increases with $Z^2$, see Eq.~\eqref{eq8}. For non-central collisions, $Z$ should be regarded as the 
	number of stopped charges, \emph{i.e.}, $Z^2 \propto N^2_{\rm part}$. On the other hand, soft hadron multiplicity is known
	to grow proportionally to  $N_{\rm part}$. The yield of bremsstrahlung photons therefore increases parametrically faster
	towards mid-rapidity than the yield of soft hadrons that decay into ``background" photons. 
\end{enumerate}
While these qualitative considerations cannot replace a realistic modeling of meson distributions and their photon decay 
kinematics, they suggest that there are experimental handles to separate bremsstrahlung photons from other ``background" sources.

\section{Conclusion}
The ALICE collaboration plans to develop a new detector (ALICE-3) with experimental acceptance in a previously uncharted, ultrasoft regime $10\, {\rm MeV} < p_T < 100\, {\rm MeV}$ and up to relatively forward pseudorapidity. As demonstrated here (Fig.~\ref{fig6}), photon bremsstrahlung due to stopping of the incoming net charge distributions is an expected phenomenon that leaves characteristic signatures in this newly accessible experimental regime. As such, it is useful for illustrating the novel opportunities of a detector design with ultrasoft acceptance. Its centrality- and $p_T$-dependence is characteristically different from that of expected backgrounds. 

Historically, the physics motivation for measuring bremsstrahlung photons is to characterize the longitudinal dynamics of stopping. As demonstrated here (Figs.~\ref{fig3} and ~\ref{fig4}), different stopping scenarios yield bremsstrahlung spectra with characteristically different pseudorapidity distribution and yield. If the pseudorapidity distribution of net charge became directly measurable with a future detector, one would have access to two experimentally challenging but complementary signatures of the same stopping phenomenon, thus allowing for much-wanted cross checks. 

Understanding bremsstrahlung photons is also of relevance in searches for imprints of other conceivable phenomena that would show up in the ultrasoft regime. For instance, unlike the situation in charged hadron distributions where Coulomb repulsion counteracts Bose-Einstein enhancement, ultrasoft $\pi^0$-yields seem ideally suited to test the quantum statistics of Cooper-Frye freeze-out distributions~\cite{Cooper:1974mv}. While we are currently not in a position to quantify these or other effects, we enjoy speculating about a future in which photon bremsstrahlung from stopping is not only measured but needs to be included in the baseline for searches of other conceivable phenomena in the ultrasoft regime. 

{\bf Acknowledgements.} We thank Federico Antinori, Maximilian Attems, Peter Braun-Munzinger, Jasmine Brewer, Adrian Dumitru, Stefan Floerchinger, Joe Kapusta, Aleksas Mazeliauskas, Johanna Stachel and Wilke van der Schee for useful questions and discussions during the preparation of this paper.


\end{document}